\documentclass[%
 preprint,
 superscriptaddress,
 amsmath,amssymb,
 aps,
 pra,
 showkeys
]{revtex4-2}

\usepackage{graphicx}
\usepackage{dcolumn}
\usepackage{bm}
\usepackage[colorlinks=true,linkcolor=black,anchorcolor=black,citecolor=black,filecolor=black,menucolor=black,runcolor=black,urlcolor=black]{hyperref}
\usepackage{physics}

\begin{document}


\title{Multi-qubit quantum state preparation enabled by topology optimization}

\author{A. Miguel-Torcal}
 \email{alberto.miguel@uam.es}
 \affiliation{Departamento de F\'isica Te\'orica de la Materia Condensada, Universidad Aut\'onoma de Madrid, E- 28049 Madrid, Spain.}
 \affiliation{Condensed Matter Physics Center (IFIMAC), Universidad Aut\'onoma de Madrid, E- 28049 Madrid, Spain.}
\author{A. Gonz\'alez-Tudela}
 \affiliation{Institute of Fundamental Physics IFF-CSIC, Calle Serrano 113b, 28006 Madrid, Spain.}%
\author{F. J. Garc\'ia-Vidal}
 \affiliation{Departamento de F\'isica Te\'orica de la Materia Condensada, Universidad Aut\'onoma de Madrid, E- 28049 Madrid, Spain.}
 \affiliation{Condensed Matter Physics Center (IFIMAC), Universidad Aut\'onoma de Madrid, E- 28049 Madrid, Spain.}
 \affiliation{Institute of High Performance Computing, Agency for Science, Technology, and Research (A*STAR), Connexis, 138632 Singapore.}
\author{A. I. Fern\'andez-Dom\'inguez}
 \email{a.fernandez-dominguez@uam.es}
 \affiliation{Departamento de F\'isica Te\'orica de la Materia Condensada, Universidad Aut\'onoma de Madrid, E- 28049 Madrid, Spain.}
 \affiliation{Condensed Matter Physics Center (IFIMAC), Universidad Aut\'onoma de Madrid, E- 28049 Madrid, Spain.}

\date{\today}

\begin{abstract} 
Using topology optimization, we inverse-design nanophotonic cavities enabling the preparation of pure states of pairs and triples of quantum emitters. Our devices involve moderate values of the dielectric constant, operate under continuous laser driving, and yield fidelities to the target (Bell and W) states approaching unity for distant qubits (several natural wavelengths apart). In the fidelity optimization procedure, our algorithm generates entanglement by maximizing the dissipative coupling between the emitters, which allows the formation of multipartite pure steady states in the driven-dissipative dynamics of the system. Our findings open the way towards the efficient and fast preparation of multiqubit quantum states with engineered features, with potential applications for nonclassical light generation, quantum simulation, and quantum sensing.
\end{abstract}

\maketitle


\section{Introduction}
The preparation and manipulation of highly entangled multiqubit states is at the core of all quantum technologies. This makes the degree of control over qubit-qubit interactions a key aspect in the assessment of material systems for their implementation. In this context, nanophotonic devices offer a wide range of strategies to tailor photon propagation and the photonic density of states at different length scales~\cite{Giannini2011,Chang2018}. This makes them stand out among other candidates for quantum hardware in terms of scalability, integration and speed of operation~\cite{Elshaari2020,GonzalezTudela2024}. For this reason, in recent years, photonic architectures structured at the nanoscale to harness photon-assisted interactions among quantum emitters (QEs, acting as qubits) have been proposed as the platform for applications such as quantum light sources~\cite{Mitsch2014,Bhaskar2017,Groiseau2024} and detectors~\cite{Najafi2015}, quantum networks~\cite{Sipahigil2016,Grim2019} and circuits~\cite{Lodahl2017,BlancoRedondo2020}, memories~\cite{Mouradian2015}, sensors~\cite{RamseySpec} or simulators~\cite{GonzalezTudela2015,Tabares2023}. 

Lately, inverse design techniques have revealed unexpected and counter-intuitive optimization pathways for nanophotonic systems~\cite{Molesky2018,So2020}. These numerical tools have proven to be particularly successful in functionalities that require a complex trade-off between conflicting mechanisms~\cite{Piggott2015,Lin2016,Lin2017}, and its impact on the field have even opened the way towards the exploration of the fundamental limits of photonic performance~\cite{Kuang2020,Chao2022} and discovery~\cite{Wiecha2021}. More recently, these numerical tools have been also exploited in the realm of quantum nano-optics, where a delicate balance between the engineering of light-matter near-field coupling and the shaping of radiation and absorption is required. Thus, inverse-designed quantum nanophotonic devices have shown a notable performance in the context of qubit entanglement formation~\cite{Liu2023,MiguelTorcal2022} and single-photon generation~\cite{Melo2023,yang2023}.

Here, we employ Topology Optimization (TO)~\cite{Bendsoe2003,Jensen2011} to inverse-design dielectric cavities that enable the preparation of Bell states~\cite{Sych2009} of QE pairs. They are obtained by using the fidelity of the density matrix of the system to these target states as the optimization function, in conditions of continuous coherent pumping and inter-emitter distances of a few natural wavelengths (comparable to the cavity dimensions). By setting the laser fields driving the emitters in phase (anti-phase), antisymmetric (symmetric) Bell-like states are obtained. Our analysis of the driven-dissipative dynamics of the QEs reveals that the small differences between the state implemented in the TO cavities and the target one translates into a slight reduction in its preparation time~\cite{Vivas2022}. Finally, we prove the versatility of our design strategy extending our investigation to QE triples and presenting a dielectric device that generates highly-entangled tripartite states with fidelities to the symmetric W state~\cite{Dur2000} comparable to those obtained from the ad-hoc optimization of the master equation parameters.     

\section{Theoretical model}
Our nanophotonic cavities are designed to host pairs and triples of distant QEs, modeled as two-level systems (with perfect quantum yield) under laser driving. The dynamics of the density matrix for the system, $\rho$, is described by a master equation of the form~\cite{Dung2002}
\begin{equation}
\mathcal{L}\rho\equiv\imath\Big[\rho,H\Big]+\sum_{i,j}\gamma_{ij}\left(\sigma_j\rho\sigma_i^{\dagger}-\frac{1}{2}\left\lbrace\sigma_i^{\dagger}\sigma_j,\rho\right\rbrace\right)=\frac{d\rho}{dt},
\label{MEq}
\end{equation}
under the assumption that the QEs are weakly coupled to their electromagnetic (EM) environment, an approximation that will be revisited below. In Eq.~\eqref{MEq}, $\sigma_i$ ($\sigma^{\dagger}_i$) is the annihilation (creation) operator for the QE labeled as $i$ (ranging from 1 to 2-3), fulfilling $\{\sigma^{\dagger}_i,\sigma_j\}=\delta_{ij}$. The Hamiltonian above can be written in the laser frame as
\begin{equation}
H=\sum_i\delta_i\sigma_i^\dagger\sigma_i+\sum_{i\neq j}
g_{ij}\sigma_i^\dagger\sigma_j+\sum_i\Omega_i(\sigma_i+\sigma_i^\dagger),
\label{HEq}
\end{equation}
where $\delta_i=\omega_i-\omega_{\rm L}$ is the detuning of the frequency of the emitters, $\omega_i$, with respect to the laser frequency, $\omega_{\rm L}$. The second term in Eq.~\eqref{HEq} reflects the coherent interaction between the QEs assisted by off-resonant EM modes with strength given by $g_{ij}$, and the last one, their laser driving with amplitudes $\Omega_i$. Finally, and back to Eq.~\eqref{MEq}, the dissipative interaction between the QEs ($i\neq j$), as well as their radiative decay ($i=j$) is also accounted for by Lindblad operators weighted by the dissipative matrix (with entries $\gamma_{ij}$), which must be positive semi-definite to ensure the physical character of the system dynamics~\cite{petruccione}.

D\"ung et al.~\cite{Dung2002} established the connection between the coherent and dissipative coupling parameters in Eqs.~\eqref{MEq}-\eqref{HEq} and the EM Dyadic Green's function~\cite{bookNovotny2012} of the dielectric environment of the QEs, obtaining $g_{i j}=\omega^{2}\vb{p}^*\mathfrak{R}\{\vb{G}(\vb{r}_{i}, \vb{r}_{j},\omega)\} \vb{p}/\hbar \varepsilon_{0} c^{2}$ and $\gamma_{i
j}=2\omega^{2}\vb{p}^*\mathfrak{I}\{\vb{G}(\vb{r}_{i},
\vb{r}_{j}, \omega)\}\vb{p}/\hbar \varepsilon_{0} c^{2}$, where $\vb{p}$ is the transition dipole moment of the QEs and $\vb{r}_{i,j}$, their position. This framework links the quantum dynamics of ensembles of identical emitters and the spatial distribution of the dielectric permittivity in their vicinity, $\epsilon(\vb r)$, which has allowed the investigation of QE entanglement generation in different nanophotonic structures~\cite{Dzsotjan2010,GonzalezTudela2011,MiguelTorcal2022}. In these works, the dyadic Green's function is evaluated at the QE frequency. We employ it here to describe emitters with slightly different natural frequencies, and detuned from the driving laser. We anticipate that the validity of this approach for the nanophotonic cavities that we obtain from the TO algorithm will be demonstrated below. In what follows, we will employ the laser frequency, $\omega_{\rm L}\simeq 2.067$ eV ($\lambda=600$ nm) for the evaluation of the $g_{ij}$ and $\gamma_{ij}$ parameters in our calculations.

The density matrix obtained from the solution of Eq.~\eqref{MEq} allows for the calculation of expectation values of any physical observable of the system. In our case, we will focus on its fidelity~\cite{Fidelity}, $F_\phi=\expval{\rho}{\phi}$, to a desired pure state, $\ket{\phi}$. This quantity ranges from 0 to 1, and we will use it as a measure of the similarity of the quantum state of the QE pair/triple in our nanophotonic system to the target one. Exploiting the dependence of the master equation parameters on the permittivity of the medium hosting the QE through the dyadic Green's function, we have developed an inverse design algorithm based on TO that provides the optimum (lossless, real-valued) dielectric map, $\epsilon(\vb{r})$, for a given target state $\ket{\phi}$ by maximizing $F_\phi$. A detailed description of the numerical method can be found elsewhere~\cite{MiguelTorcal2022}, we only sketch it here. Inspired by recent reports~\cite{Mignuzzi2019,Bennett2020,Bennett2021}, it consists of a nested iterative procedure in which $\epsilon(\vb r) $ is shaped with the precision of the spatial discretization used to solve Maxwell's Equations. Starting from free space, in each iteration, the effect of a small, local increment of the permittivity, $\delta\epsilon$, on the target function is assessed for each mesh element. By keeping only those that contribute to enlarge $F_\phi$,  dielectric cavities with optimum performance (for a set of given constraints) are attained. The high speed and efficiency of the algorithm resides in the use of first-order Born scattering series and the exploitation of Lorentz reciprocity in the evaluation of the effect of the local dielectric variations on the Dyadic Green's functions~\cite{bookNovotny2012}. 

\begin{figure}[tb!]
\centering
\includegraphics[width=0.7\linewidth]{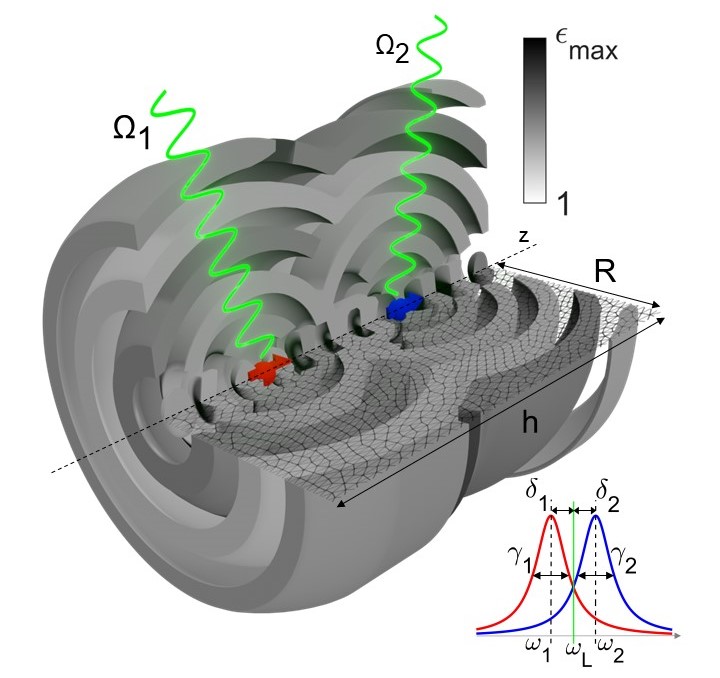} 
\caption{TO-designed nanophotonic cavity, of radius $R$ and height $h$, hosting a QE pair. The emitters are aligned and oriented along $z$-direction, and driven coherently by laser fields of amplitude $\Omega_i$ (wavy green lines). The light grey mesh renders the spatial discretization in the solution of Maxwell's Equations, and the permittivity map is coded from white (vacuum) to black ($\epsilon=\epsilon_{\rm max}$). The bottom-right panel illustrates the spectral position and Lorentzian-type lineshapes of the QEs, with their linewidths and detunings with respect to the laser frequency $\omega_{\rm L}$.}
\label{fig:1}
\end{figure}
Figure \ref{fig:1} illustrates the TO method and an inverse-designed cavity hosting a QE pair. The algorithm is interfaced with the finite-element EM solver implemented in Comsol Multiphysics$^{\rm TM}$, whose spatial discretization is sketched by the dark gray thin mesh. The QEs are separated a distance $d_{12}$ along $z$-direction, with their dipole moments parallel to it. This emitter configuration allows us to exploit the azimuthal symmetry of the system to solve Maxwell's Equations within the $rz$-plane only. As a  result, we obtain cylindrical cavities with rotational symmetry, radius $R$ and height $h$. The dielectric function varies from 1 (white) to its maximum, $\epsilon_{\rm max}$ (black), which varies from one design to another. We set a threshold, $\epsilon_{\rm max}\leq 9$, corresponding to semiconductor materials such as GaP~\cite{Cambiasso2017} in the visible range, to remain in the typical parameter regime of nanophotonics. In the bottom right corner, the lineshape of the two emitters is represented, with natural frequencies $\omega_i$ and linewidths $\gamma_i=\gamma_{ii}$. Both are detuned from the laser frequency $\omega_{\rm L}$.

\section{Results}
First, we design dielectric cavities to prepare distant QE pairs into maximally-entangled pure states in the steady-state regime, $\mathcal{L}\rho=0$ in Eq.~\eqref{MEq}. We set the dimensions of the structures to $R=6.25\lambda$ and $h=16.67\lambda$ ($\lambda=600$ nm), and the dipole moment of the emitters to $|\vb{p}|=1\,e\cdot$nm, which yields a free-space decay rate $\gamma_0=\omega^3\abs{\vb{p}}^2/3\pi\hbar\epsilon_0c^3=2.3\, \mu$eV. We choose the even and odd Bell states, 
\begin{equation}
\ket{\rm{+\pm}}=\frac{1}{\sqrt{2}}\big[\ket{ge}\pm\ket{eg}\big],
\label{Bell}
\end{equation}
as our target, with $g$ ($e$) indicating the ground (excited) state of each QE, and generate two different sets of devices, resulting from the maximization of the corresponding fidelities, $F_{+\pm}$. In accordance with recent literature~\cite{Pichler2015}, we make the  detunings of the QE frequencies symmetric with respect to the laser frequency, and significantly smaller than their linewidth, $|\delta_i|/\gamma_0=0.2$ ($i=1,2$). The laser pumping strengths are set to $|\Omega_i|/\gamma_0=0.7$, and their parity is given by the symmetry of the target state, having $\Omega_1=\mp\Omega_2$ for $\ket{\rm{+\pm}}$. The diffraction limit of classical optics imposes a lower bound for the inter-emitter distance, $d_{12}\gtrsim\lambda/2$, to make a reliable control over the two laser fields possible. 

\begin{figure}[tb!]
\centering
\includegraphics[width=0.7\linewidth]{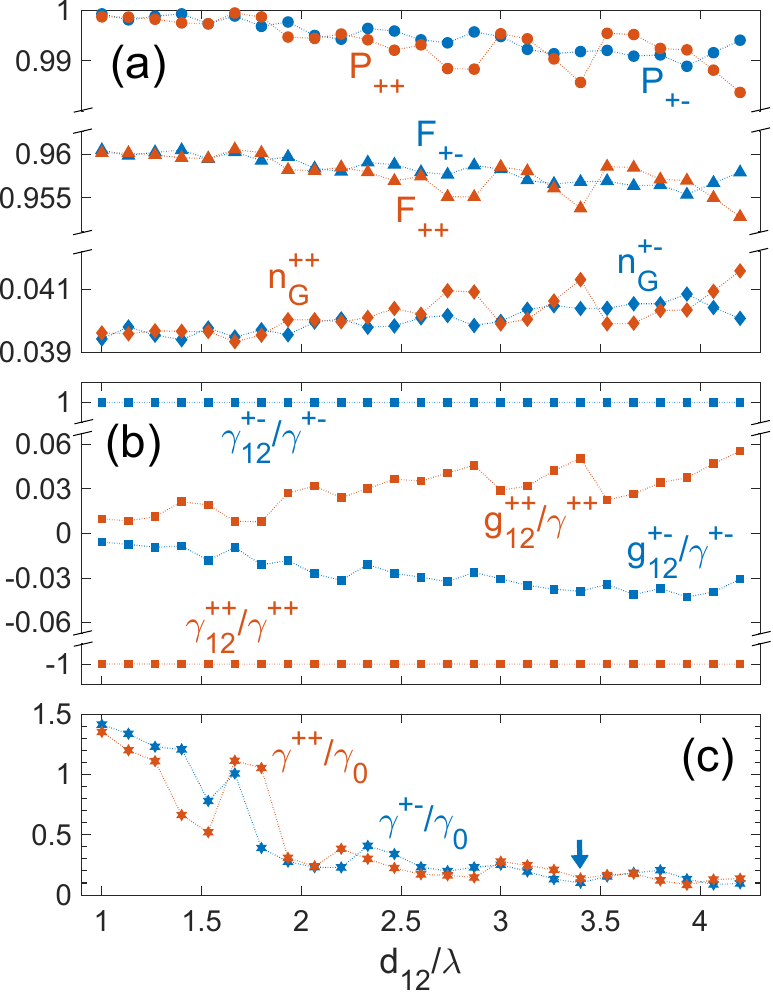}
\caption{(a) Fidelity to even and odd Bell states, $F_{+\pm}$, versus inter-emitter distance for TO cavities with $R=6.25\lambda$ and $h=16.67\lambda$. Purity, $F_{+\pm}$, and ground state population, $n_G^{+\pm}$, of the attained states. The detuning and pumping parameters are $|\delta_i|/\gamma_0=0.2$ and $|\Omega_i/|\gamma_0=0.7$. (b) Dissipative, $\gamma_{12}$, and coherent, $g_{12}$, coupling strengths normalized to the QEs decay rate $\gamma=\sqrt{\gamma_1\gamma_2}$ as a function of $d_{12}$ for same devices as panel (a). (c) Purcell factor, $\gamma/\gamma_0$, versus emitter-emitter distance for all the cases above. The blue arrow indicates the configuration considered in Figures~\ref{fig:3} and ~\ref{fig:3_2}.}
\label{fig:2}
\end{figure}
Figure~\ref{fig:2} analyzes the performance of the inverse-designed cavities obtained for inter-emitter distances between 1 and 4$\lambda$, and targeting even (in orange dots) and odd (in blue dots) Bell states. In Figure~\ref{fig:2}(a), the fidelities (used as optimization functions) are shown in connected triangles. We obtain $F_{+\pm}>0.95$ for all emitter-emitter distances and both symmetries. The small deviation from unity is caused by the finite size of the devices, which we restrict to the micron scale to remain in the domain of nanophotonics technology. This is manifested in the slight decreasing trend of the fidelity as a function of $d_{12}$ for both sets of data, and the fact that, in some cases, the TO algorithm was terminated because the maximum permittivity condition, $\epsilon_{\rm max}=9$, was reached in some position within the device.   

To clarify the nature of the quantum states sustained by the TO cavities, their purity is plotted in connected circles in Figure~\ref{fig:2}(a). This is calculated as $P_{+\pm}=\Tr{\rho_{+\pm}^2}$, where $\rho_{+\pm}$ is the system density matrix (the subscripts indicate the target Bell state). We can observe that it is above 0.98 in all cases, indicating the pure character of the states formed in the devices~\cite{Nielsen_Chuang2010}. Note as well that the purities present a decreasing slope very similar to $F_{+\pm}$. The deviation of $\rho_{+\pm}$ from the Bell states in Eq.~\eqref{Bell} becomes clearer by computing the ground state populations, $n_G^{+\pm}=\expval{\rho_{+\pm}}{gg}$. They are rendered in rhombuses in Figure~\ref{fig:2}(a), obtaining $n_G^{+\pm}\sim0.04$ and a positive slope with increasing distance. Again, this indicates that the radiation losses experienced by the TO cavities due to their micron-sized dimensions are behind the failure to obtain $\ket{\rm{+\pm}}$ with higher accuracy.

Once we have verified the capability of the cavities to produce  highly-entangled steady states for QEs several laser wavelengths apart, we explore the physical mechanism behind their operation. For this purpose, we plot in Figure~\ref{fig:2}(b) the coherent and dissipative coupling strengths they realize as a function of $d_{12}$. Both are normalized to the collective decay rate of the system, defined as $\gamma=\sqrt{\gamma_1\gamma_2}$~\cite{Ficek2002}. We can observe that the dissipative coupling is maximized~\cite{Carminati2019}, reaching absolute values equal to this collective decay, $\gamma_{12}^{+\pm}/\gamma^{+\pm}=1$, and its sign is positive (negative) for odd (even) target Bell states . On the contrary, $g_{12}^{+\pm}$ acquires vanishing values, with opposite sign to $\gamma_{12}^{+\pm}$. These values tend to reproduce the master equation parameters previously reported in theoretical proposals for emitter entanglement through cavity or waveguide dissipation~\cite{Alharbi2010,Kastoryano2011,Zheng2013,Pichler2015,vivasviana2023,Ramos14,Ramos16,plenio99}. In all these proposals, however, the free-space emission was only considered phenomenology as an extra parameter, but no realistic calculation was performed. In the proposal with plasmonic waveguides~\cite{GonzalezTudela2011}, such factor was taken into account showing how their ability to suppress far-field emission offer a feasible realization for this dissipative entanglement mechanism. However, this came at the expense of high absorption losses, which effectively restricts the QE-QE distances to the sub-wavelength regime. To our knowledge, our TO dielectric cavities are the first platform that implement efficiently this phenomenon for distant emitters thanks to the ability to optimize both the generation of long-range interactions and the suppression of the far field emission at the same time.

Our TO algorithm is able to produce dielectric cavities that operate efficiently within a wide range of inter-emitter distances. In order to keep the ratio $\gamma_{12}^{+\pm}/\gamma^{+\pm}$ in its maximum value for all $d_{12}$, they must modulate the QE radiation into free-space in different ways. This is shown in Figure \ref{fig:2}(c), which plots the Purcell factor,  $\gamma^{+\pm}/\gamma_0$ for all the cavities in the panels above. At small distances, the dissipative coupling is strong, and the emission rate of the QEs is Purcell-enhanced (by a factor 1.5) to achieve a maximal fidelity. On the contrary, for large $d_{12}$, the QE interactions are weak, requiring a strong reduction in their decay rate and yielding $\gamma^{+\pm}/\gamma_0\simeq0.1$. As anticipated, the devices showcase a 15-fold difference in the Purcell factor, which is key for the formation of dissipatively-entangled quantum states with similar $F_{+\pm}$ for QEs separated by very different distances.
\begin{figure}[tb!]
\centering
\includegraphics[width=0.7\linewidth]{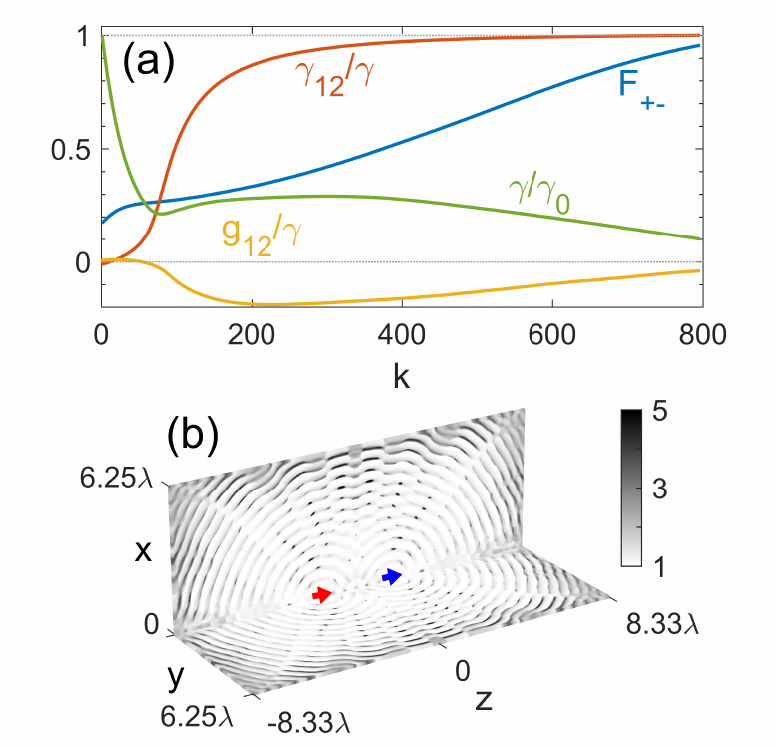}
\caption{(a) Fidelity, normalized coupling strengths and Purcell factor versus iteration step for the TO cavity targeting the odd Bell state for $d_{12}=3.4\lambda$. (b) Permittivity map obtained as a result of the TO procedure in panel (a). The dielectric constant is represented in grey scale between 1 (white) and 5 (black). The red and blue arrows indicate the position and orientation of the QEs.}
\label{fig:3}
\end{figure}

Having assessed the performance of the TO cavities for Bell state preparation, we focus next on different aspects of the inverse-design procedure. To do so, we select a particular device, indicated by the blue vertical arrow in Figure \ref{fig:2}(c) ($d_{12}=3.4\lambda$, odd symmetry), and use it as a test bed for the inspection of the operation of our optimization algorithm. Figure \ref{fig:3}(a) plots the fidelity to the target state (blue line), the coupling strengths (red and yellow lines) and the collective Purcell factor (green) versus the iteration step $k$. Each step corresponds to a complete scan within $rz$-section of the cavity, evaluating the Green's function and consolidating or discarding a $\delta\epsilon=0.003$ increment in each mesh point. Three different regimes can be distinguished in the evolution of these quantities along the iterative procedure towards the condition $F_{+-}=1$. First, a sharp drop in the Purcell factor takes place, while both couplings grow slowly in absolute value and different sign. Between the steps 100 and 300, $\gamma/\gamma_0$ varies very little, while $\gamma_{12}/\gamma$ increases quickly, approaching its maximum and $g_{12}$ reaches its minimum value. For $k>300$, both the Purcell factor and coherent coupling tend smoothly to 0, while the dissipative coupling, already very close to the condition $\gamma_{12}/\gamma=1$, converges towards it. In all this process, the fidelity grows almost linearly from 0.2 at $k=0$ to $F_{+-}=0.96$ at $k=800$, when the cavity implements master equation parameters very similar to those previously reported for the formation of dissipatively entangled, dark states: $\gamma_{12}=\gamma$, $g_{12}=0$.

The dielectric map, $\epsilon(\vb{r})$, of the nanophotonic cavity obtained at the end of the optimization procedure described above is shown in Figure~\ref{fig:3}(b). The permittivity is fully characterized within the $rz$-plane, but to facilitate its visibility, it is displayed within $xz$- and $yz$-planes. The QEs are sketched as blue and red arrows along the $z$-axis. The grey scale codes the dielectric constant linearly from 1 (white) to $\epsilon_{\rm max}=5$ (black). Two different regions can be distinguished in $\epsilon(\vb{r})$. Few wavelengths apart from the QEs and near the edges of the cavity, elliptical-shaped, high-contrast periodic reflectors are apparent over a smooth $\epsilon\simeq3$ background. These we can link to the reduction of the collective Purcell factor that minimizes the radiation decay experienced by the emitters. In their near-field and centered around them, two sets of lower-contrast, spherical-shaped shells can be observed. These are embedded into a $\epsilon\simeq1$ background. These elements contribute to the tailoring of the emitter-emitter interactions, maximizing (vanishing) its dissipative (coherent) coupling. Small, deeply-subwavelength and isolated rings of high permittivity are distributed along the radial direction in between these two regions. Note that the maximum permittivity in the device acquires only a moderate value, well below the threshold set for the TO algorithm. 

\begin{figure}[tb!]
\centering
\includegraphics[width=0.7\linewidth]{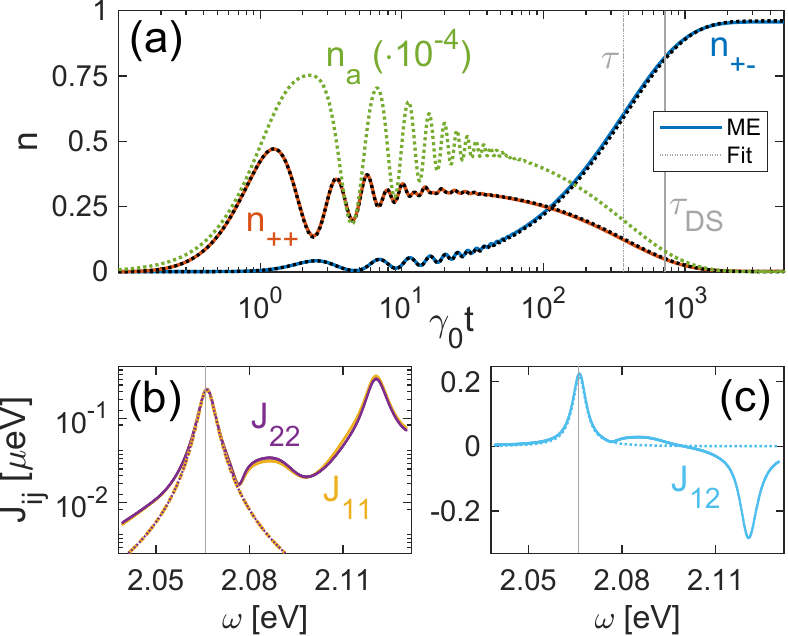}
\caption{(a) Population dynamics of the QE pair within the TO cavity in Figure~\ref{fig:3}(b) under even coherent driving and initially in their ground state, $n_G(0)=1$. The transients obtained from Eq.~\eqref{MEq} and Eq.~\eqref{MEq2} are rendered in colour solid and black dotted lines, respectively. Vertical grey lines indicate the preparation times of the TO steady-state, $\tau$, and the phenomenological pure dark state, $\tau_{\rm DS}$. (b) and (c) show the spectral density at the emitters position (yellow, purple) and the cross spectral density (light blue). Dotted lines render their single-mode fitting at the laser frequency (vertical line).}
\label{fig:3_2}
\end{figure}
Figure~\ref{fig:3_2}(a) explores the preparation time (from the onset of the laser driving) required for the emergence of the steady-state in the cavity in Figure~\ref{fig:3}(b). It plots the population dynamics in the first excitation manifold and in the Bell state basis, $n_{+\pm}(t)=\expval{\rho(t)}{+\pm}$. The system is initially in its ground state, $n_{G}(0)=1$. We can clearly observe that once the QE pair is pumped, the population is transferred first into the even Bell state, and $n_{++}(t)$ (orange) develops a plateau (preceded by significant oscillations) that showcases a meta-stable regime~\cite{Vivas2022} that extends up to $t\simeq40\gamma_0^{-1}$. In this time window, $n_{+-}(t)$ is negligible. At longer times, the population of the odd Bell states grows quickly, reaching $n_{+-}\simeq0.96$ at $t\gtrsim10^3\gamma_0^{-1}$, which sets the preparation time for the cavity steady-state. To obtain another estimation of this time, we compute the inverse of the Liouvillian gap of Equation~\eqref{MEq}~\cite{manz2,Albert2014}. The value obtained, $\tau=350\gamma_0^{-1}$, is indicated by the vertical grey dotted line in Figure~\ref{fig:3_2}(a). We can benchmarck this estimation against the inverse of the Liouvillian gap for the phenomenological master equation that yield pure dark Bell states in Ref.~\cite{Vivas2022},  $\tau_{DS}=700\gamma_0^{-1}$ (see vertical grey solid line). Thus, we can conclude that the small deviation from the target state in the TO cavities allows for a shorter preparation time than the exact Bell state generated under an ad-hoc theoretical model.  

To gain insight into the population dynamics in Figure~\ref{fig:3_2}(a), we investigate next the role of the cavity fields as intermediaries of the QE-QE interactions behind them. To do so, we must refine our model of the system to account for the photonic degrees of freedom in its quantum density matrix, $\varrho$. Thus, we perform EM calculations for the spectral densities, $J_{ij}(\omega)=\gamma_{ij}(\omega)/2\pi$~\cite{Medina2021} for each emitter and between them, shown in Figure~\ref{fig:3_2}(b) and (c), respectively. In these panels, the solid lines render the EM simulations for our TO device, and dotted lines their single Lorentzian fittings~\cite{SanchezBarquilla2022}
\begin{equation}
J_{ij}(\omega)=\frac{\mathcal{G}_i \mathcal{G}_j}{\pi}\frac{\Gamma_{a}/2}{(\omega-\omega_{a})^2+(\Gamma_{a}/2)^2} \label{J}
\end{equation}
in the vicinity of the laser frequency (indicated by vertical grey lines). The fitting parameters, $\mathcal{G}_1\simeq\mathcal{G}_2=7.75\gamma_0=17.1\,\mu$eV and $\Gamma_a=2459\gamma_0=5.4$ meV, yield very good agreement with the full spectra within a 0.03 eV window around $\omega_{\rm L}$ (much larger than the QE detunings). 

Equation~\eqref{J} allows for the direct parametrization of a master equation of the form~\cite{Medina2021}
\begin{equation}
\mathcal{L'}\varrho\equiv\imath\Big[\rho,H'\Big]+\Gamma_a\left(a \varrho a^{\dagger}-\frac{1}{2}\left\lbrace a^{\dagger} a,\varrho\right\rbrace\right)+\sum_i\gamma_0\left(\sigma_i \varrho \sigma_i^{\dagger}-\frac{1}{2}\left\lbrace \sigma_i^{\dagger} \sigma_i,\varrho\right\rbrace\right)=\frac{d\varrho}{dt},
\label{MEq2}
\end{equation}
with $H'=(\omega_a-\omega_{\rm L})a^{\dagger}a+\sum_i\delta_i\sigma_i^\dagger\sigma_i+
\mathcal{G}_{i}\sigma_i^\dagger a+\Omega_i\sigma^{\dagger}_i + h.c.$ It describes the coupling of both emitters to their EM environment, with strengths $\mathcal{G}_{i}$, approximated by a single cavity mode with frequency $\omega_a$ and linewidth $\Gamma_a$~\cite{Li2016}. Black dotted lines in Figure~\ref{fig:3_2}(a) plot the QE populations obtained with this refined model. The excellent agreement with the transients in solid lines prove the accuracy of the original description in the weak-coupling, quasi-degenerate regime ($|\delta_i|=0.2\gamma_0\simeq2\gamma$) in which the inverse-designed cavity operates~\cite{Mccauley2020}. Moreover, the solution to Equation~\eqref{MEq2} enables us to compute the population transient for the cavity mode, $n_{a}(t)={\rm tr}\{a^{\dagger}a\varrho(t)\}$. This is rendered in green dotted line in Figure~\ref{fig:3_2}(a), and shows that the plateau in $n_{++}(t)$ coincides with the time window in which the cavity population is non-negligible. Thus, the cavity mode sustains this meta-stable regime in the quantum dynamics, beyond which, its population decays and the two QEs, already in the target Bell state, become effectively decoupled from their EM environment.

Up to here, we have designed nanophotonic cavities to host quantum states of QE pairs, but our TO algorithm can be applied to any emitter ensemble. As discussed above, the limiting factor is the calculation of the EM Dyadic Green's function, which is greatly simplified if the QEs are aligned. Thus, to prove the versatility of our method we consider now QE triples, located along $z$-axis and with dipole moments parallel to it. There exist multiple ways to generate entanglement in tripartite systems, which have been the object of much research in recent years~\cite{Buchleitner2004,Brunner2012}. We focus on a well-known class of three-qubit states, the so-called $W$ states~\cite{Acin2001,Freudenthal2009}. These are a class of pure states that present high robustness against noise and losses, which means that they retain the maximum possible amount of bipartite entanglement when any one of the three QEs is lost (traced out)~\cite{Cirac2000}. In particular, we take the symmetric W state
\begin{equation}
\ket{+++}=\frac{1}{\sqrt{3}}\big[\ket{gge}+\ket{geg}+\ket{egg}\big], \label{W}
\end{equation}
as the target for our TO algorithm. We keep the same cavity dimensions ($R=6.25\lambda$ and $h=16.67\lambda$) as before which, to accommodate a third QE, requires a reduction of the inter-emitter distances to $d_{12}=d_{23}=d=1.17\lambda$ (where the extremal QEs are labelled as 1 and 3, and the central one as 2).    

To proceed with the cavity design maximizing $F_{+++}$, the fidelity to the state in Equation~\eqref{W}, we must first set the external parameters (driving amplitudes and emitter detunings) of the Liuovillian. In the case of QE pairs, these were set in accordance with recent literature~\cite{Pichler2015}. We do not have such analytical insight in QE triples, and need to use a different approach. Operating at the master equation level, we perform a particle-swarm-optimization (PSO)~\cite{PSO2} of its parameters, taking $F_{+++}$ as the objective function. Imposing invariance under the exchange of QEs 1 and 3, 10 quantities remain to be optimized: 6 internal ones, describing the emitter-emitter interactions and decay rates, and 4 external ones. To limit the range of parameter values, we used the EM Dyadic Green's function for a  bulk medium with $\epsilon_{\rm max}=9$ (threshold permittivity in the TO algorithm), to estimate the spatial variation they can experience within the inter-emitter distance, $d$. This way, we found that the conditions ${\rm sign}\{g_{12,23}\}=-{\rm sign}\{g_{13}\}$ and ${\rm sign}\{\gamma_{12,23}\}=-{\rm sign}\{\gamma_{13}\}$ had to be fulfilled. Our PSO computation in the constrained 10-dimensional parameter space involved $2\cdot10^3$ particles, up to $5\cdot10^3$ iteration steps, and 1000 runs under different initializations. The quantum steady-state obtained this way presents a fidelity $F_{+++}=0.91$ and a purity $P=0.99$, indicating that it corresponds to a pure state of the QE triple  slightly different from  Equation~\eqref{W}. This PSO procedure does not only allow us to benchmark the performance of our TO algorithm, it also provides us with the 4 external parameters that it requires: the QE-laser detunings, $\delta_{1,3}/\gamma_0=0.55$ and $\delta_2/\gamma_0=-0.3$, and laser amplitudes, $\Omega_{1,3}/\gamma_0=0.33$ and $\Omega_2/\gamma_0=-0.73$. 

\begin{figure}[tb!]
\centering
\includegraphics[width=0.6\linewidth]{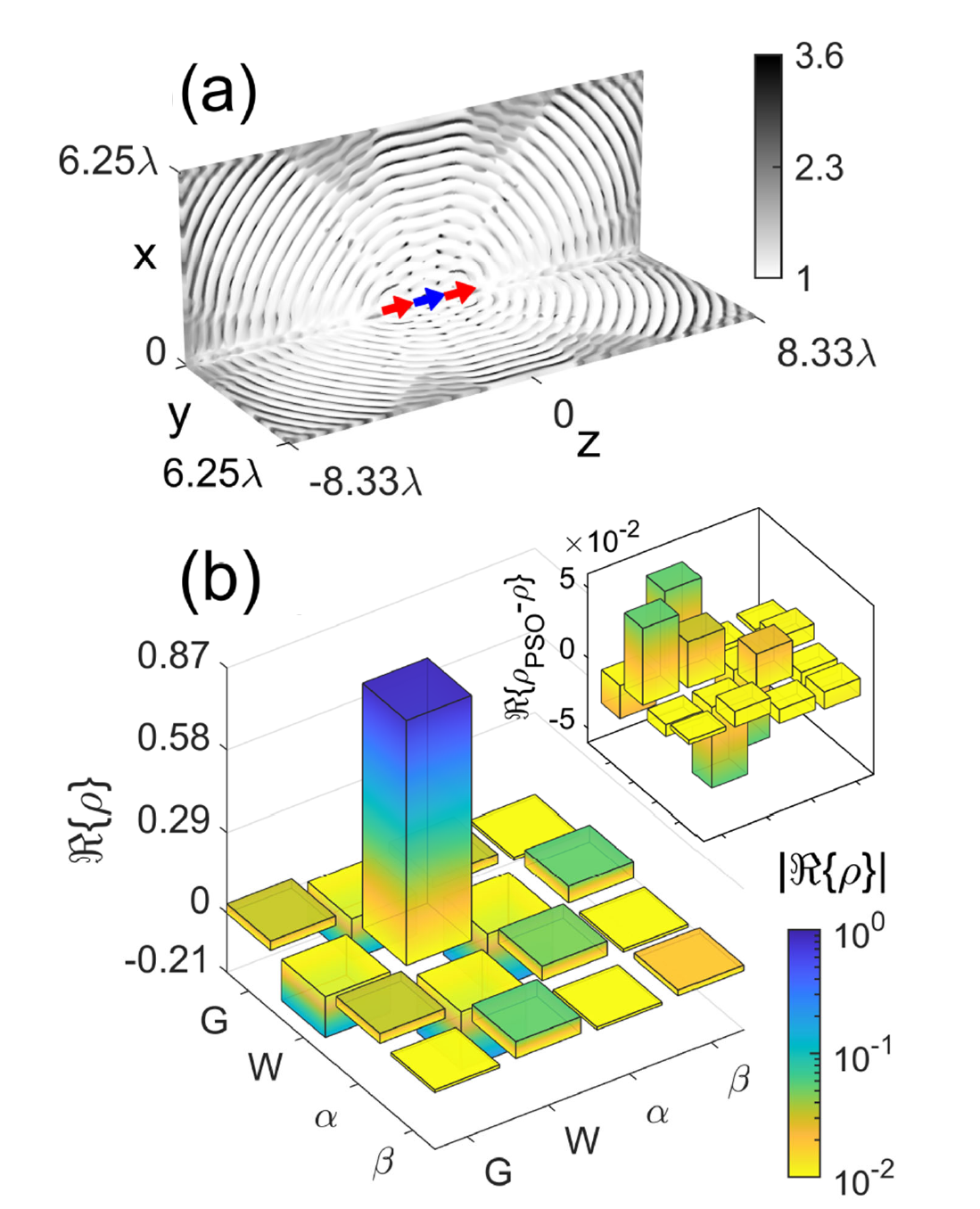}
\caption{(a) Permittivity map for the inverse-designed cavity maximizing the  fidelity to the W state in Equation~\eqref{W} for an inter-emitter distance $d=1.17\lambda$, and the same dimensions as the device in Figure~\ref{fig:3}(b). The dielectric constant is coded in white-to-black linear scale. The position and orientation of the emitters along $z$-direction is indicated by red and blue arrows. (b) Quantum tomography of the steady-state density matrix (real part) for a QE triple in the cavity in panel (a). The colors represent the absolute value of the real part of the density matrix in logarithmic scale. Inset: difference between the TO density matrix and a phenomenological one, resulting from the direct particle-swarm-optimization of the master equation parameters maximizing $F_{+++}$ (same color scale as in the main panel).}
\label{fig:4}
\end{figure}
Figure \ref{fig:4}(a) shows the map $\epsilon(\vb{r})$ for the TO cavity maximizing $F_{+++}$. Its structure resembles very much that in Figure~\ref{fig:3}(b), presenting periodic reflectors with high permittivity contrast and a larger, $\epsilon\sim2$, background near its boundaries than at the central region, where $\epsilon\sim1$. Remarkably, $\epsilon_{\rm max}=3.6$, lower than that in Figure~\ref{fig:3}(b). This originates from the shorter inter-emitter distance, as the cavities for Bell state preparation at $d_{12}\simeq d=1.17\lambda$ (not shown) present a permittivity range very similar to that in Figure \ref{fig:4}(a). The main panel of Figure~\ref{fig:4}(b) shows the tomography of the steady-state density matrix hosted by the dielectric cavity in panel (a). It is restricted to the ground and the single excitation manifolds, where the basis is formed by $\ket{+++}$ and the states $\ket{\alpha}=\tfrac{1}{\sqrt{3}}[\ket{gge}-a\ket{geg}-b\ket{egg}]$ and $\ket{\beta}=\tfrac{1}{\sqrt{3}}[\ket{gge}-b\ket{geg}-a\ket{egg}]$, with $a=(1+\sqrt{3})/2$ and $b=(1-\sqrt{3})/2$. The vertical axis displays the real part of population and coherences, and the color scale codes its absolute value in logarithmic scale. By simple inspection, we can extract $F_{+++}=\expval{\rho}{+++}=0.87$, whose discrepancy ($\sim0.04$) with the one for the PSO Liouvillian is very similar to the deviation of $F_{+\pm}$ from unity in Figure~\eqref{fig:2}(a). Thus, we can conclude that the performance of the cavities is similar in both cases. To further verify the similarity between the quantum state in the TO device and its PSO counterpart, in the inset of Figure~\ref{fig:4}(a), we present the tomography of the difference between the two density matrices. The deviations are most apparent in the W-state population and its coherences with the ground state, and these remain in the same range, below 0.05. 

Finally, we analyze in more detail the nature of the QE-triple state implemented by the nanophotonic cavity. It presents a large purity, $P=0.94$, and the single excitation section of the tomography in Figure~\ref{fig:4}(b) indicates that the state amplitudes in the bare basis are not equal, contrary to $\ket{+++}$. The calculation of the populations in this basis yield $\expval{\rho}{gge}=0.210$, $\expval{\rho}{geg}=0.527$ and $\expval{\rho}{egg}=0.204$, revealing that weight of QE 2 is larger than the two extremal ones. As discussed above, the degree of entanglement of  W states exhibit a strong robustness against the disposal of one of the qubits. Indeed, the bipartite states that result from the tracing out of one of the QEs, $\rho^{(k)}=\Tr_k\{{\rho}\}$ (where $k=1,2,3$), in Equation~\eqref{W} present a significant Wootters concurrence~\cite{Wootters1998}, $C(\rho^{(k)})=2/3=0.667$ for all $k$. Note that $C=1$ for a maximally-entangled two-qubit state, such as the Bell states in Equation~\eqref{Bell}. The concurrence calculation for the partial traces of the density matrix in Figure~\ref{fig:4}(b) yields $C(\rho^{(1)})=0.624$, $C(\rho^{(2)})=0.344$, $C(\rho^{(3)})=0.632$. Thus, the degree of bipartite entanglement obtained by tracing out the extremal QEs is similar to that in the perfectly symmetric W state. However, it is slightly lower if the intermediate QE is lost. This evidences the higher sensitivity to decoherence effects in the intermediate emitter of the quantum state in the inverse-designed TO cavity. 

\section{Conclusion}
To conclude, we have generated pure quantum steady states of QE pairs and triples under coherent driving conditions through the inverse-design of their dielectric environment. By means of a topology-optimization algorithm that acts at the level of the electromagnetic Dyadic Green’s function, we have obtained nanophotonic cavities that engineer simultaneously the coherent and dissipative interactions between the emitters and their radiative decay. First, we have performed a thorough study of the capability of our devices to prepare even and odd Bell states, showing that they exploit a dissipation-driven mechanism to entangle pairs of quantum emitters separated several natural wavelengths apart. Analyzing the population dynamics in the system, we have shown that the small discrepancy between the pure states hosted by the cavities and exact Bell states translates into shorter preparation times.  Finally, we have tested the versatility of our approach by applying it to a triple of quantum emitters, successfully realizing a highly entangled tripartite state akin to a symmetric W state. We believe that our results prove that inverse design is a powerful tool for the conception, implementation and refinement of quantum-optical hardware based on nanophotonic platforms, with direct application in areas such as quantum sensing and simulation.  

\subsection*{Funding}
This work has been sponsored the Comunidad de Madrid through the 2020 CAM Synergy Project Y2020/TCS-6545 (NanoQuCo-CM) and the  Projects TED2021-130552B-C21 and TED2021-130552B-C22 funded by MICIU/AEI/10.13039/501100011033 and by FEDER Una manera de hacer Europa. AIFD also acknowledges funding from the Spansh Project PID2021-126964OB-I00 and from the European Union's Horizon Europe Research and Innovation Programme under agreement 101070700. AGT thanks the CSIC Research Platform on Quantum Technologies PTI-001, the Spanish project PID2021-127968NB-I00, and a 2022 Leonardo Grant for Researchers and Cultural Creators, and BBVA Foundation. AMT, FJGV and AIFD thank the support from MICIU through the Mar\'ia de Maeztu Programme for Units of Excellence in R\&D (CEX2023-001316-M). 

\subsection*{Acknowledgements}
We thank Alejandro Vivas-Via\~na and Carlos S\'anchez Mu\~noz  for fruitful discussions.


\bibliography{main}

\end{document}